\documentclass[fleqn,usenatbib]{mnras}

\usepackage[T1]{fontenc}

\DeclareRobustCommand{\VAN}[3]{#2}
\let\VANthebibliography\thebibliography
\def\thebibliography{\DeclareRobustCommand{\VAN}[3]{##3}\VANthebibliography}

\usepackage{graphicx}	
\usepackage{amsmath}	
\usepackage{amssymb}	

\usepackage{newtxtext,newtxmath}

\title[3D MHD simulations of the SNR  CTB~109]{3D MHD simulations of the supernova remnant CTB~109}

\author[Castellanos-Ram\'irez et al.]{
A. Castellanos-Ram\'irez,$^{1}$\thanks{E-mail: acastellanos@astro.unam.mx (ACR)}
P. F. Vel\'azquez,$^{2}$
J. Cant\'o$^{1}$
\\
$^{1}$Universidad Nacional Aut\'onoma de M\'exico, Instituto Astronom\'\i a, Ap. 70-264, CDMX, 04510, M\'exico\\
$^{2}$Instituto de Ciencias Nucleares, Universidad Nacional Aut\'onoma de M\'exico, Ap. 70-543, CDMX, 04510, M\'exico
}

\date{Accepted XXX. Received YYY; in original form ZZZ}

\pubyear{2021}

\begin{document}
\label{firstpage}
\pagerange{\pageref{firstpage}--\pageref{lastpage}}
\maketitle

\begin{abstract}
We examine the evolution and emission of the supernova remnant (SNR) CTB109 using three-dimensional magneto-hydrodynamics simulations. The SNR evolves in a medium divided by a plane interface into two media with different densities and pressure equilibrium. Our results reveal that a remnant with the characteristics of CTB109 is formed provided the supernova (SN) explosion takes place in the less dense medium and also if the interstellar magnetic field (ISMF) is almost uniform. Finally, we conclude that the quasiparallel mechanism can explain the brightness synchrotron emission and the position angle of the projected ISMF reported in previous works.
\end{abstract}

\begin{keywords}
MHD -- polarization -- radiation mechanisms: general -- ISM: supernova remnants -- ISM: individual objects (CTB109) -- methods: numerical -- 
\end{keywords}



\section{Introduction}
\subsection{Generalities}
A supernova remnant (SNR) is the result of the interaction of material ejected by a supernova explosion (SN), at the end of the life of a massive star, and its surrounding environment. The SNRs are the main producers of some heavy elements, and both these and the elements and dust of the surroundings are injected into the interstellar medium, altering its composition and dynamics.

CTB~109 (SNR G109.1-1.0) is a supernova remnant with semicircular morphology. This remnant was first discovered in X-ray by \citet{Gregory1980} through observations with the Einstein satellite, and later by \citet{Hughes1981} in radio ($\lambda$49 cm) with the Westerbork Synthesis Radio Telescope. In both cases, the observations show a similar morphology: a well-defined semicircular shell with roughly the same radius to the east, and an incomplete structure to the west. This remnant was cataloged as a shell-type SNR by \citet{Downes1983}. Even though this SNR hosts the anomalous X-ray pulsar (AXP) 1E 2259+586 (first detected by \citealt{Gregory1980}, and listed as an X-ray pulsar by \citealt{Fahlman1981}), according to data obtaining with the ROSAT Position Sensitive Proportional Counter (PSPC) and Broad Band X-Ray Telescope (BBXRT) by \citet{Rho1997} and  EPIC data of XMM-Newton reported by \citet{Sasaki2004}, the spectrum of the remnant is completely thermal.

The semicircular shape observed in CTB~109 has been verified in numerous works both in X-ray (see e.g. \citealt{Rho1993,Hurford1995,Parmar1998,Sasaki2004,Sasaki2006,Sasaki2013, Nakano2015}) and in radio (see e.g. \citealt{Hughes1984,Braun1986,Tatematsu1987a,Kothes2002,Kothes2006,Tian2010,Sun2011}) and has been explained first using CO observations by \citet{Tatematsu1987a}. These data show that the western part of the SNR has encountered a giant molecular cloud (GMC). Notably, the GMC extends even in front of the remnant (an armlike CO ridge in the north part of the remnant). However, \citet{Hughes1981} did not detect significant radio emission in the west part of the remnant. Also, \citet{Sasaki2004} did not detect X-ray emission in this same region. All of these data confirm that the SNR has hit the GMC and has been practically stopped.

In addition to the GMC, there are two notable phenomena lodged in CTB~109: On the one hand, as we mentioned before, there is the anomalous X-ray pulsar (AXP) 1E 2259+586 with no radio emission detection (\citealt{Hughes1984}). On the other hand, the so-called lobe; is an extremely bright region in X-rays. It was discovered by\citet{Tatematsu1987b}. There is currently evidence that the high X-ray emission produced by the lobe may be due to the interaction between the SNR shock and a dense cloud (see \citealt{Sasaki2004}, \citealt{Sasaki2013}). Both works highlight the fact that the X-ray spectrum in the lobe is completely thermal. 

\subsection{Distance and age of CTB~109}
Determining the distance to CTB~109 has been an arduous process. The interested reader can consult the works of \citet{Kothes2012} and \citet{Sanchez2018}, and the references therein for a broader discussion about the complicated process of calculating the distance. Here we mention only three of these works (in chronological order), whose value coincides with the most accepted value distance to CTB ~ 109.

\begin{itemize}
\item \citet{Kothes2002}, compare the dynamic of the HI and CO in the environment surrounding the SNR and compare them with spectroscopic distances and radial velocities of nearby HII regions. From this analysis, they obtain that CTB~109 is located at a distance of $3.0\pm 0.5$ kpc.

\item \citet{Kothes2012} computed the distance to CTB~109 analyzing the interaction of the SNR with the GMC, also they use the data of molecular and neutral material towards the remnant and finally used the HI velocity profiles. They found a value of $3.2\pm 0.2$ kpc. This is, currently, the most accepted value for the distance to CTB~109.

\item \citet{Sanchez2018} estimated a distance to CTB~109 using the velocity–distance diagram of \citet{Kothes2012} through the study of the radial velocity of the optical filaments in the remnant. Further, they considered the distance to the  AXP 1E 2259+586. With these, they determined a value of $3.1\pm 0.2$ kpc.
\end{itemize}

Regarding the age of CTB~109, \citet{Sasaki2004} used the distance obtained by \citet{Kothes2002} ($3.0\pm 0.5$ kpc) and the Sedov-Taylor solution applied only to the XMM-Newton data to the eastern side of the remnant (which is the farthest from the pulsar and has a lower density than the rest of the remnant), they determined that the age of CTB~109 is about $(8.8\pm 0.9)\times 10^3$ yr. They also obtained the radius of the remnant $R=16\pm 1$ pc; an initial energy of the explosion $E_0=(7.4\pm 2.9)\times 10^{50}$ erg, and, using the jump shock conditions, a preshock density of $n_0=(0.16\pm 0.02)$ cm$^{-3}$. Analogously, \citet{Sasaki2013} used the data obtained with Chandra (for a pair of outer regions in the remnant), obtaining an age of $(14\pm 2)\times 10^3$ yr.

On the other hand, using observational data of the Susaku X-ray Imaging Spectrometer, and a distance of $3.2\pm 0.2$ kpc obtained by \citet{Kothes2012} (together with  the angular size of $\sim 16^{\prime}$), \citet{Nakano2015} obtained a radius of $R=16\pm 1$ pc. Moreover, with the Sedov-Taylor solution, and the strong shock case for the shock front velocity of the SNR, they estimated an age of $(14\pm 2)\times 10^3$ yr, and a preshock density in the range of $n_0=0.1-0.3$ cm$^{-3}$. As we can see, this is the same age as reported by \citet{Sasaki2013}.

In a recent work, \citet{Sanchez2018} used the kinematic parameters derived in their work, together with the model developed by \citet{McKee1975}, obtained and age of $9.0\times 10^{3}$ yr for the Sedov-Taylor phase (this value is a little lower than that obtained in other works, but similar to that reported by \citealt{Sasaki2004}). They also computed an initial energy deposited by the explosion of  $E_0=1.8\times 10^{50}$ erg.

\subsection{Radio emission, polarization and spectral index}
There are plenty of works about the radio emission of CTB ~ 109. \citet{Downes1983} made radio observations of CTB~109 with the Effelsberg 100-m telescope at 2.7 GHz. They measured the Stokes parameters $I$, $Q$, and $U$. For the first time, they clearly identified two bright peaks located at the northeast and south side of the remnant. On the other hand, they reported that they did not detect linearly polarized radio emission.

 Radio observations of CTB~109 at $\lambda$49 cm, $\lambda$21 cm using the Westerbork Synthesis Radio Telescope, and at $\lambda$4.6 cm with the Synthesis Telescope of the Dominion Radio Astrophysical Observatory were made by \citet{Hughes1984}. Using a data set containing both theirs and others, they found an spectral index $\alpha=0.5\pm0.004$\footnote{Throughout this text, we will use the convention that the flux density is given
by $S_{\nu}\propto \nu^{-\alpha}$, where $\alpha$ is the spectral index and $\nu$ the frequency.}.

Based on a polarization study through observations at 10 GHz with the Nobeyama 45-m telescope, \citet{Tatematsu1987a} measured polarized intensity and the degree of polarization around CTB~109. Also, they detected linear polarization for first time. According to their results, the polarized intensity is in a range from 10 to 20 mJy/beam. The degree of polarization is $\ge$ 30 \%. Regarding the polarization orientation, the position angle lies between 110$^{\circ}$ and 160$^{\circ}$. They did not find evidence of a tangential magnetic field. The position angle of the magnetic field is in the range between 20$^{\circ}$ and 70$^{\circ}$.

Recently, \citet{Kothes2002} conducted 408 and 1420 MHz radio continuum observations of CTB~109 (and polarization in the latter) using the Synthesis Telescope of the Dominion Radio Astrophysical Observatory (DRAO). They showed that the surrounding cold material and the radio emission from the remnant are closely related. The emission shows a narrow inner shell surrounded by a wide external shell and ends right in the region where both shells are incomplete(the northwest region). The brightest regions are in the northeast and south (where both shells meet). In between, some thin regions are emitting in radio, which gives an idea of the inhomogeneous structure of the surrounding environment.

Additionally, \citet{Kothes2006} presented a catalogue of SNRs (included CTB~109) observed with the Canadian Galactic Plane Survey. They measured flux densities, spectral indices and fractional polarization. Acording to the observations, they detected weak polarization from the remnant. As well, they obtained an spectral index $\alpha=0.5\pm0.04$.

In the same way, \citet{Sun2011} carried out a study of spectral and polarization properties of 51 small SNRs (including CTB~109), using observations of the Sino-German $\lambda$6 cm polarization survey of the Galactic plane. They estimated the integrated polarization flux density (showing polarization images) and the average polarization percentage. Also, they obtained a new spectral index for CTB~109, $\alpha=0.45\pm0.04$, based on their new observations.

In this work we model the synchrotron emission of CTB 109 using MHD simulations. This paper is organized as follows:
in section 2 we present previous numerical studies of this remnant and the initial setup of our simulacions; section 3 explains the calculation of the radio and thermal X-ray emission, section 4 presents and discusses the obtained results, finally, in section 5, we summarize our conclusions.

\section{Numerical models}

\begin{figure*}
	\includegraphics[width=14cm]{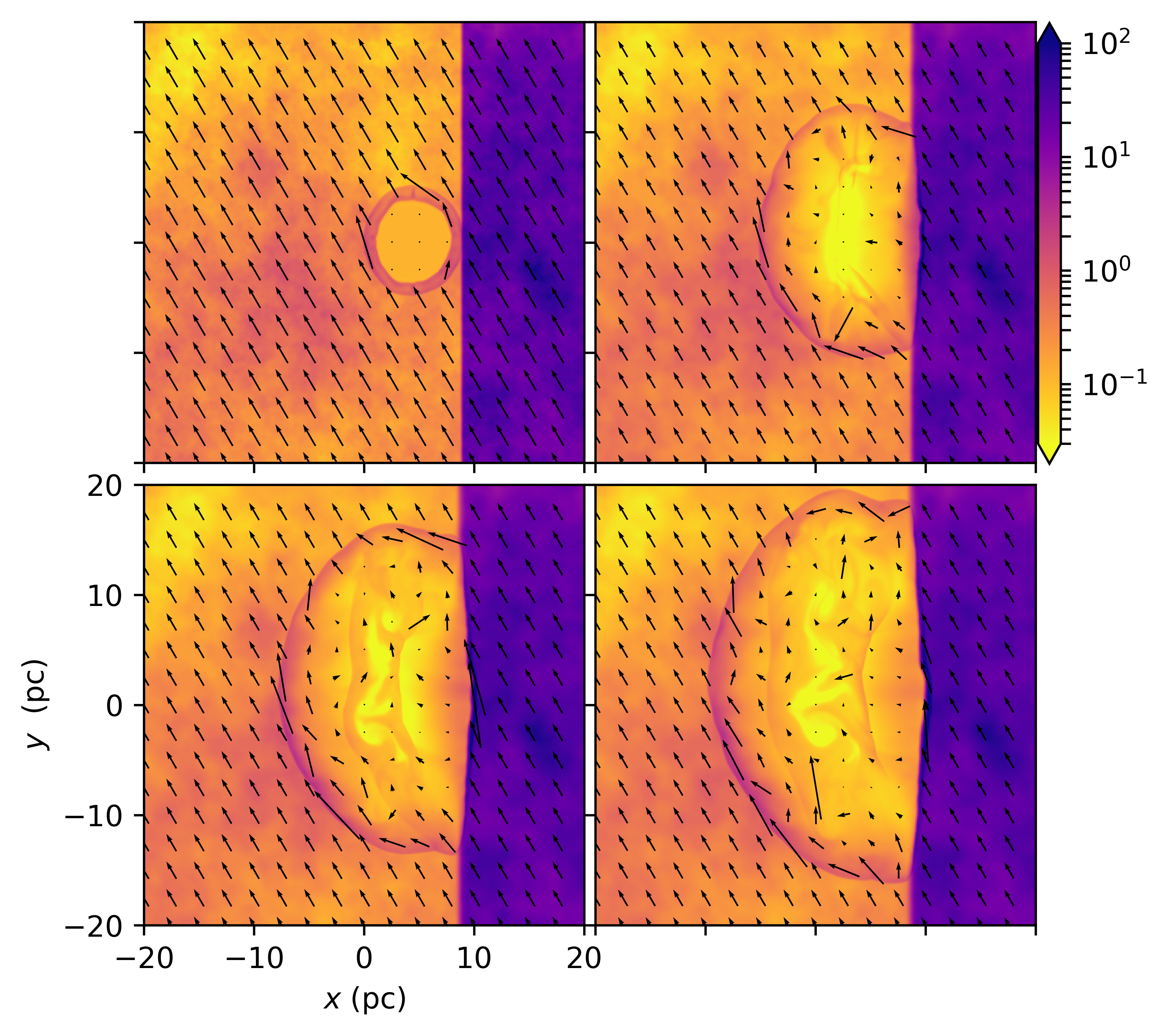}
    \caption{Density stratification map on the $xy$ plane at four different times: top-left, 1000 yr; top-right, 5000 yr; bottom-left, 9000 yr, and bottom-right, 13000 yr. The density values are shown with the logarithmic colour
scale given by the right bar (in cm$^{-3}$). The black arrows represent the magnetic field vectors. The top-left panel shows the initial distributions of the vector field (with magnitude $B_0=5.0\times 10^{-6}$ G and making an angle of $120^{\circ}$ with the $x$ axis, except in the shock region). The rest of the panels show the evolution of the magnetic field around the shock. The size of the arrows are related to the magnitude of $\mathbf{B}$.}
    \label{fig:Fig1}
\end{figure*}

\subsection{Previous works}
\citet{Wang1992} use the thin shell approximation (see \citealt{Kompaneets1960}) to model the morphology of CTB~109. They solve the hydrodynamics equations in spherical symmetry dividing the computational domain into symmetrical rings. Their model assumes an initial energy $E_0=3.6\times 10^{50}$ erg and a constant density of the medium $n_0=0.13$ cm$^{-3}$, a density $n_{c,0}=36$ cm$^{-3}$ for the cloud, and a distance between the supernova explosion an the interface of 2 pc. With these initial conditions, they obtain an age of $1.3\times 10^4$ yr, which is in good agreement with those reported in the observations.

Furthermore, they performed calculations to obtain the X-ray and radio emission. For the first case, they obtain the temperature based on the shock jump conditions. Their results agree well with those reported by \citet{Gregory1980}. For the radio emission, they followed the prescription of \citet{Chevalier1982}. The nonthermal radiation emitting by CTB~109 is in form of synchrotron radiation from a power-law distribution of relativistic electrons and assuming equipartition between the density energy and the magnetic field. The flux obtained from their calculations is a bit low compared to the flux observed by \citet{Hughes1984}.

\citet{Castro2012} developed spherically symmetrical hydrodynamical models to study the broadband characteristics of CTB~109. These models include efficient cosmic-ray acceleration, nonthermal emission, and a self-consistent calculation of the X-ray thermal emission. Their main idea is to use the broadband properties of the remnant to explain a MeV–GeV emission source detected in $\gamma-$ray. They considered a uniform circumstellar medium with $n_0=0.5$ cm$^{-3}$, a magnetic field $B_0=4.5$ $\mu$G, and a deposited initial energy $E_0=10^{51}$ erg. They obtained an age of $1.1\times 10^4$ yr (adopting a distance of 2.8 kpc). For this model, they consider the X-ray emission as a mix between thermal and nonthermal. As a result, to obtain the best fit for the thermal X-ray emission, their model considered that both leptons and hadrons contribute to the emission of the $\gamma$-ray source.

\citet{Bolte2015} performed 3D hydrodynamical simulations to explore the nature of CTB~109, in particular, they were interested in explaining the bright emission produced in the Lobe, leaving aside the interaction with the GMC. \citet{Bolte2015} introduce an inhomogeneous medium with more realistic conditions since they derived these initial conditions directly from the CO emission data observations from the Canadian Galactic Plane Survey (CGPS). They explore models based on two cases: The XMM-Newton observations of \citet{Sasaki2004}, and Chandra observations (\citealt{Sasaki2013}). For XMM-Newton, they chose an initial environment density $n_0=0.155$ cm$^{-3}$, while for Chandra, they chose an initial environment density $n_0=0.3$ cm$^{-3}$. In both cases, they imposed the supernova explosion with an initial energy $E_0=10^{51}$ erg. The results showed that the age of the remnant in the case of XMM-Newton is about $8\times 10^3$ yr and for Chandra, $1.1\times 10^4$ yr, comparable values to those obtained by the observations. 

They also compare their computed X-ray emissivities with those obtained by \citet{Sasaki2004} for the Lobe, and conclude that the numerical computations are in good agreement with the observations. On the other hand, although their work did not include magnetic fields, they conclude that their model could represent the remnant's structure. This is based on the fact that the higher shock velocities are expected in the regions where the electrons are more efficiently accelerated, and this is related to a high synchrotron emission. Because of this, \citet{Bolte2015} show that the regions with high velocities are distributed in a shell-like structure in the north-east having a higher concentration in the south, which agrees with the observations showed by \citet{Kothes2002}.

On the other hand, in recent work, \citet{Meyer2021} presented 2D MHD simulations of the impact of the ISM magnetic field along with different phases in the stellar wind that affects the environment surrounding the supernova explosion and varying the velocity of the progenitor in each of the phases. They found that the morphology of the remnant is molded with the circumstellar medium created by previous stellar stages.

Therefore, according to \citet{Meyer2021} CTB~109 is surrounded by a circumstellar medium produced by the wind of a red-supergiant. Inside this medium, the supernova is interacting with a ring structure left by the Wolf-Rayet in the previous stage of evolution. Nevertheless, they do not consider the impact of the presence of the molecular cloud in the expansion of the supernova remnant.

\subsection{The numerical setup}

We performed our numerical models using the {\sc guacho} code (\citealt{Esquivel2009, Villarreal2018}). This code solves the idealized MHD equations of the mass, momentum, and energy conservation coupled with magnetic field induction on a three-dimensional, Cartesian fixed mesh. The energy density is given by $E=\rho v^2/2+\epsilon + B^2/(8\pi)$, being $\rho$, $v$, and $B$ the gas density, the velocity magnitude and the intensity of the magnetic field, respectively. The internal energy is  $\epsilon=P/(\gamma-1)$, i.e., the ideal gas equation, where $P$ is the pressure and $\gamma$ is the specific heat ratio, which is set as $5/3$.
The code includes radiative cooling as a parametrized function of the temperature. We use the cooling function described in \citet{Dalgarno1972}. Thereby, the cooling curve is tabulated for temperatures above $10^4$ K assuming the gas is fully ionized. Below this value, the cooling is turned off. On the other hand, above $10^8$ K, we follow a $T^{1/2}$ law suitable for the free-free regime. Our study does not consider the energy loss on the SNR shock front due to the relativistic electron acceleration.

The computational domain is a cube of 40 pc on a side with a total of 512$^3$ points in the computational mesh. 

The environment where the SN explosion takes place consists of two media, which are in pressure equilibrium and separated by the plane $x=8.5$~pc. In the region $x<8.5$~pc, we impose a medium with a number density $n_0=0.3$~cm$^{-3}$ \citep{Sanchez2018} and a temperature $T_0$ equal to 5000~K. For $x\geq 8.5$~pc, we have the GMC, which is one hundred times heavier \citep[i.e., 30~cm$^{-3}$,][]{Sasaki2013,Bolte2015} with a temperature of 50 K.

The supernova explosion was located at $(x,y,z)=(4.5,0,0)$~pc, with an initial energy $E_0=4.0\times 10^{50}$ erg and an ejected mass of 3 M$_{\odot}$, which is uniformly distributed into a sphere of radius $R_0\approx1.1$ pc. Of this energy, 95\% is kinetic, and the remaining is thermal. The velocity increases linearly with $r$, until finally reaching the value $v_0=1.3\times 10^{9} \sqrt{f_{ek}E_{51}/M_{*}}$~cm s$^{-1}$ at $r=R_0$; where $f_{ek}$ is the fraction of the kinetic energy, $E_{51}$ is $E_0$ in units of 10$^{51}$~erg, and $M_{*}$ is the ejected mass given above. From our parameters, we found that $v_0\approx 4600$ km s$^{-1}$.

As we have seen before, there are several features in the observations that reveal a highly inhomogeneous environment in which the expansion of the SNR developed. Thereby, to construct the initial conditions that simulate an inhomogeneous medium, we follow the work of \citet{Esquivel2003} to generate a noisy density structure consistent with a turbulent interstellar medium. For this environment, we modulate the density by a fractal structure with a spectral index of 11/3. Therefore, we imposed 20\% density perturbations on the whole computational domain but keeping the pressure equilibrium condition.

Finally, we consider a uniform component of the interstellar magnetic field (ISMF) $\mathbf{B}=B_0(\cos\beta\hat{x}+\sin\beta\hat{y})$, i,e, placed such that $\beta=120^{\circ}$ with the $x$ axis (in the $xy$ plane, and therefore, making an angle of 30$^{\circ}$ with the interface between the medium and the GMC) throughout all the computational domain. The magnitude of this field is $B_0=5.0\times 10^{-6}$ G, which it is a typical value for the Galactic magnetic field.

\section{Synthetic Emission Maps}
\begin{figure*}
\begin{center}
\includegraphics[width=16cm]{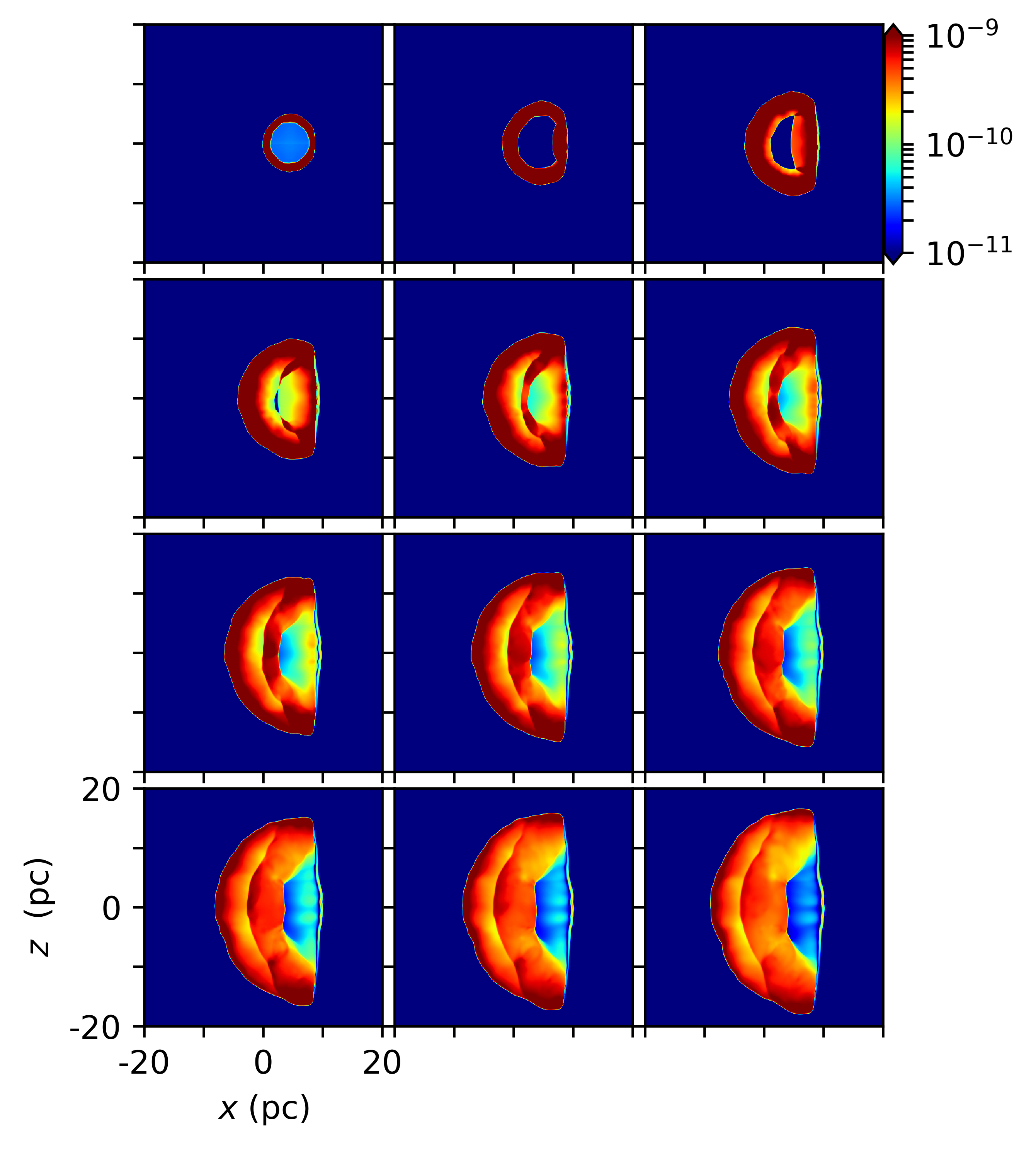}
\caption{Temporal evolution for pressure on the $xz$-plane from 1000 yr (top-left) to 12000 yr (bottom right). The panels are displayed every 1000 yr. The pressure values are shown with the logarithmic colour
scale given by the right bar (in dyn cm$^{-2}$)}
\label{fig:Fig2}
\end{center} 
\end{figure*}

\subsection{Radio synthetic maps}
In order to explore the radio emission from our model, we computed synthetic synchrotron intensity maps from our simulations. For each cell of our computational grid we compute the synchrotron emissivity. We will use the Stokes parameters $I$, $Q$, and $U$ for this purpose. The calculation is done as explained below (the interested reader can also review the works of \citealt{Schneiter2015,Cecere2016,Velazquez2017,Aroche2020}). 

Following the work of \citet{Jun1996}, the synchrotron specific intensity for each point $(x,y,z)$ in the computational domain can write as:
\begin{equation}
j_s(x,y,z,\nu)=\kappa P^{2\alpha}\rho^{1-2\alpha} B^{\alpha+1}_{\perp}\nu^{-\alpha}
\label{eq:js}
\end{equation}  
where $P$ and $\rho$ are the pressure and density of the gas, $\nu$ is the observed frequency \citep[10~Ghz,][]{Tatematsu1987a}, $B_{\perp}$ is the component of the magnetic field perpendicular to the line of sight (LoS), and $\alpha$ is the spectral index \citep[set as 0.45,][]{Sun2011}.
 
The $\kappa$ parameter contains information about the acceleration mechanism of relativistic electrons: the quasi-parallel, where we have $\kappa \propto \cos^2(\theta_B)$  being $\theta_B$ the obliquity angle (the angle between the shock normal and the post-shock magnetic field); and the quasi-perpendicular, with $\kappa \propto \sin{^2(\theta_B)}$ \citep[see][]{Orlando07,Aroche2020}. 

The total intensity of the synchrotron emission gives us our synthetic synchrotron emission maps. This can obtain as (see \citealt{Cecere2016}):
\begin{equation}
I(x,y,\nu)=\int_{LoS}j_s(x,y,z,\nu)dz.
\label{eq:I}
\end{equation}  

On the other hand, we can compute the Stokes parameters $Q$ and $U$ from the specific intensity as:
\begin{equation}
Q(x,y,\nu)=\int_{LoS}f_p\,j_s(x,y,z,\nu)\cos(2\phi)dz,
\label{eq:Q}
\end{equation}  
\begin{equation}
U(x,y,\nu)=\int_{LoS}f_p\,j_s(x,y,z,\nu)\sin(2\phi)dz,
\label{eq:U}
\end{equation}
with $\phi$ the position angle of the local magnetic field in the plane of the sky, and  $f_p$ is the degree of linear polarization, related to the spectral index $\alpha$ by,
\begin{equation}
f_p=\frac{\alpha+1}{\alpha+5/3}.
\end{equation}  

Once the parameters $Q$ and $U$ have been calculated, we find that the linearly polarized intensity is:
\begin{equation}
I_p(x,y,\nu)=\sqrt{Q^2(x,y,\nu)+U^2(x,y,\nu)},
\end{equation}
and the magnetic field position angle (PA) is obtained by,
\begin{equation}
    \Phi_B(x,y,\nu)=\frac{1}{2}\arctan\bigg( \frac{U(x,y,\nu)}{Q(x,y,\nu)}\bigg)
\end{equation}

\subsection{X-ray synthetic maps}

We estimate the X-ray luminosity from our hydrodynamical simulations in order to obtain synthetic thermal X-ray maps and compare with observations. 

We have also calculated synthetic thermal X-ray maps similarly as \citet{Castellanos-Ramirez2015}. We assume the low-density regime and use the ionization equilibrium of \citet{Mazzotta1998}. Thus, the X-ray emission coefficient $j_\nu$ is given as a function of the density and temperature such that $j_\nu(n,T)=n^2_e\xi(T)$, where $n_e$ is the electron density, $T$ the temperature computed from our numerical simulations (under the assumption of the electronic and ionic temperature is the same), and $\xi(T)$ is a smooth function exclusively of temperature. This function was computed for the range [0.35-8] keV and standard solar metallicity using the CHIANTI atomic database (\citealt{Dere1997,Landi2006A}), adding the effects of interstellar absorption assuming a column density of $N_H=6\times10^{21}$ cm$^{-2}$.
We also computed the X-ray emission from synthetic maps for the case without interstellar absorption. Finally, we integrated the X-ray emission coefficient over the specific photon energy range. It is important to note that the ionization equilibrium may not be the real ionization state in the plasma, as shown by \citealt{Hughes1985}, who found that taking ionization equilibrium underestimate the X-ray emission in the case of energies below 2 keV.

\section{Results}

\begin{figure*}
	\includegraphics[width=14cm]{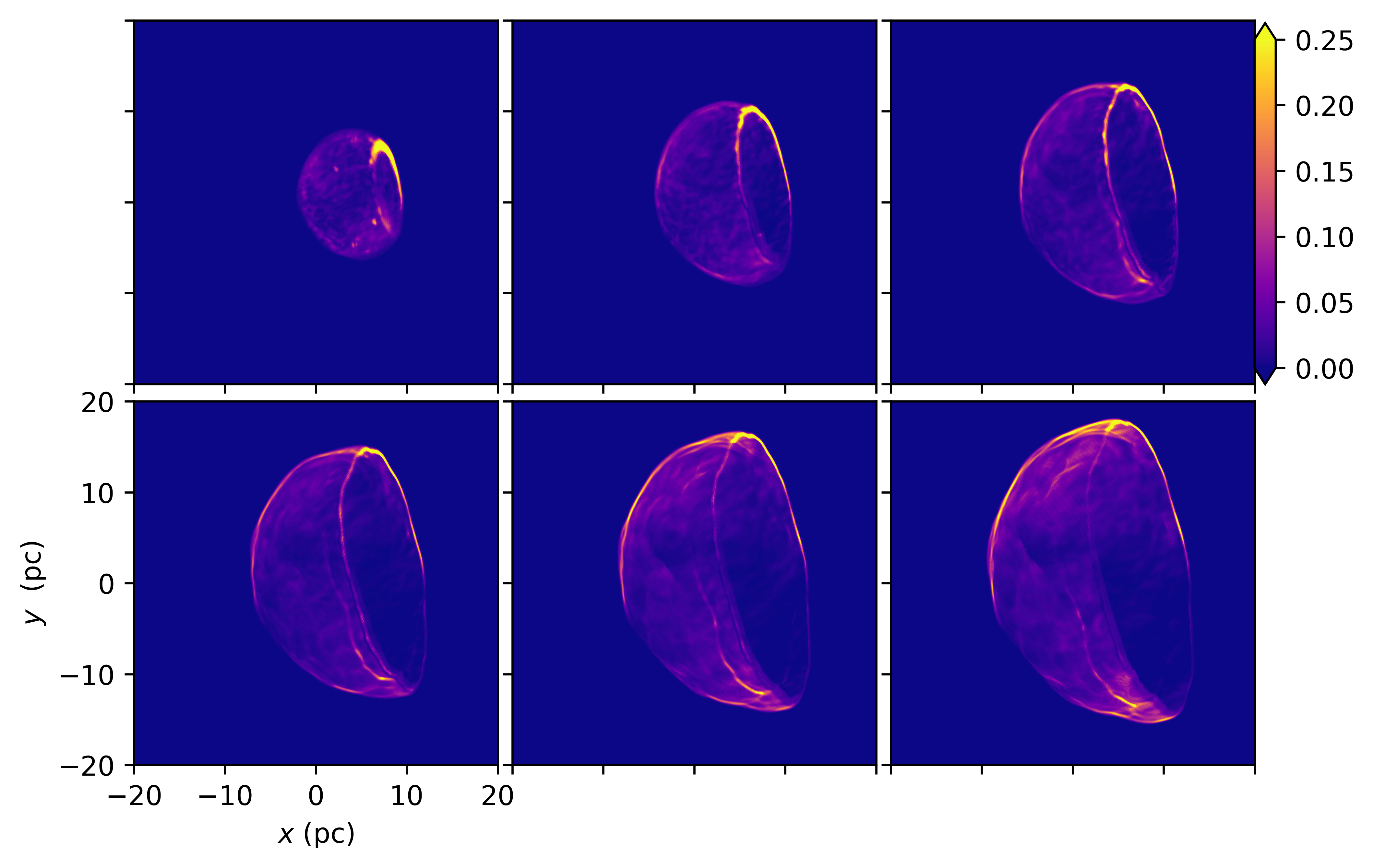}
    \caption{Synthetic map of the Stokes parameter $I$ for the case of a rotation of 30 degrees around the x-axis and 20 degrees on the y-axis and for a temporal evolution from 2000 to 12000 yr. The time between panels is 2000 yr. We can see that the size of the numerical model in the last panel closely approximates that reported observationally, $\sim$ 16 pc. The linear colour
scale given by the right bar indicates the synchrotron flux in arbitrary units.}
    \label{fig:Fig3}
\end{figure*}

\subsection{Temporal density and pressure evolution}

Figure \ref{fig:Fig1} displays the density distribution on the $xy$-plane for four integration times. We also show the magnetic field vectors over the entire computational mesh. The top left panel displays the collision between the SNR shock front and the cloud, which is just beginning at an integration time of 1000 yr. The remaining panels show further evolution after the collision. On the one hand, we observe, as expected, that the explosion has been stopped almost entirely by the dense cloud. On the other hand, although the ISM density is turbulent, the SNR shock front practically conserves the semispherical morphology.

In these last cases, we observe that the perturbation favours a clumpy structure formed within the remnant. This morphology seems to be due to the bounce of the shock once it has found the cloud. Finally, we also see the behaviour of the magnetic field inside the SNR at the edge of the shock. As we can notice, the direction and magnitude of the magnetic field are affected by the presence of the two shock structure in the SNR.

We also observe that at $t=13000$ yr, the structure formed by the remnant is very close to the edge of the computational box (bottom right panel of Figure ~\ref{fig:Fig1}). Therefore, we take $t=12000$ yr as the best time that reproduces the observational conditions in CTB ~ 109.

Figure \ref{fig:Fig2} presents pressure maps for studying the collision in more detail on the $xz$-plane. This figure shows the temporal evolution, each 1000 yr, from 1000 to 12000 yr. The upper-central panel of this figure corresponds to the time (2000 yr) when the encounter between the right part of the SNR shock front and the dense cloud has already occurred. Then it is observed how this part of the shock front is reflected on the left, starting to sweep up the SNR material. Thus, a double shock feature forms as time passes, as we observe on the following maps. As we will explain in more detail later, this double shock front will affect the synchrotron emission.

\subsection{Synthetic synchrotron and X-ray emission maps}

For comparing with observations, we performed several synthetic synchrotron and X-ray emission maps. We made these synthetic maps after rotating our computational domain $30^{\circ}$ around $x$ and $20^{\circ}$ around $y$ axes (in our original $xyz$ system,  $xy$ is the plane of the sky and the $z$-axis the LoS). We obtain the Stokes parameters (Eqs.: \ref{eq:I}-\ref{eq:U}) and the X-ray emission integrating their emissivities along the LoS.

\citet{Tatematsu1987a} obtained the synchrotron emission (at 10 GHz) and the PA distribution of CTB109. They found that the direction of the magnetic field is almost orthogonal to the bright shock front in the Northeastern region of the remnant. Similar results were reported by \citet{Reynoso2013} for SN 1006 and \citet{moranchel2017} for the SNR G296.5+10. These two last works conclude that the quasiparallel mechanism explain both the synchrotron morphology and the PA distribution, which was confirmed  by \citet{Aroche2020}.
Based on these previous works, we think that the quasiparallel mechanism is suitable for explaining the observed morphology of SNR CTB109. 

Figure~\ref{fig:Fig3} shows the synthetic synchrotron images (obtained for the quasiparallel acceleration mechanism) from 2000 to 12000 yr of evolution. At 2000 yr, we see strong radio emission produced by the collision with the dense cloud. In general, the remnant exhibits a semispherical shape to the left for all times, which corresponds to the principal shock front. All maps have minor variations in their brightness distribution. These variations are due to the SNR shock wave evolves into a turbulent medium. Notably, the upper half of this shock is brighter than the lower half region. As a first approach to understand this brightness distribution,  we note that the synchrotron emission depends strongly on $B_{\perp}$ and weakly on $P$ and $\rho$ (see Eq. \ref{eq:js}). Besides, we consider the quasi-parallel acceleration mechanism, which gives a polar cap morphology for a remnant expanding into an almost uniform ISMF. Thus, the increase in the synchrotron emission of the upper half region results from the ISMF orientation and the quasi-parallel acceleration mechanism. 

At 6000 yr (top-right panel, Figure~\ref{fig:Fig3}), secondary arc-like structure, fainter than the main shock front, has developed next to the cloud. This feature moves to the left in the following panels, resembling the double arch feature observed in the CTB109 remnant. As we observe in the pressure maps (see Fig.~\ref{fig:Fig2}), this structure is nothing more than the part of the shock front reflected after colliding with the cloud. Because the remnant does not expand inside the cloud, no synchrotron emission is coming from this region.

\begin{figure*}
	\includegraphics[width=14cm]{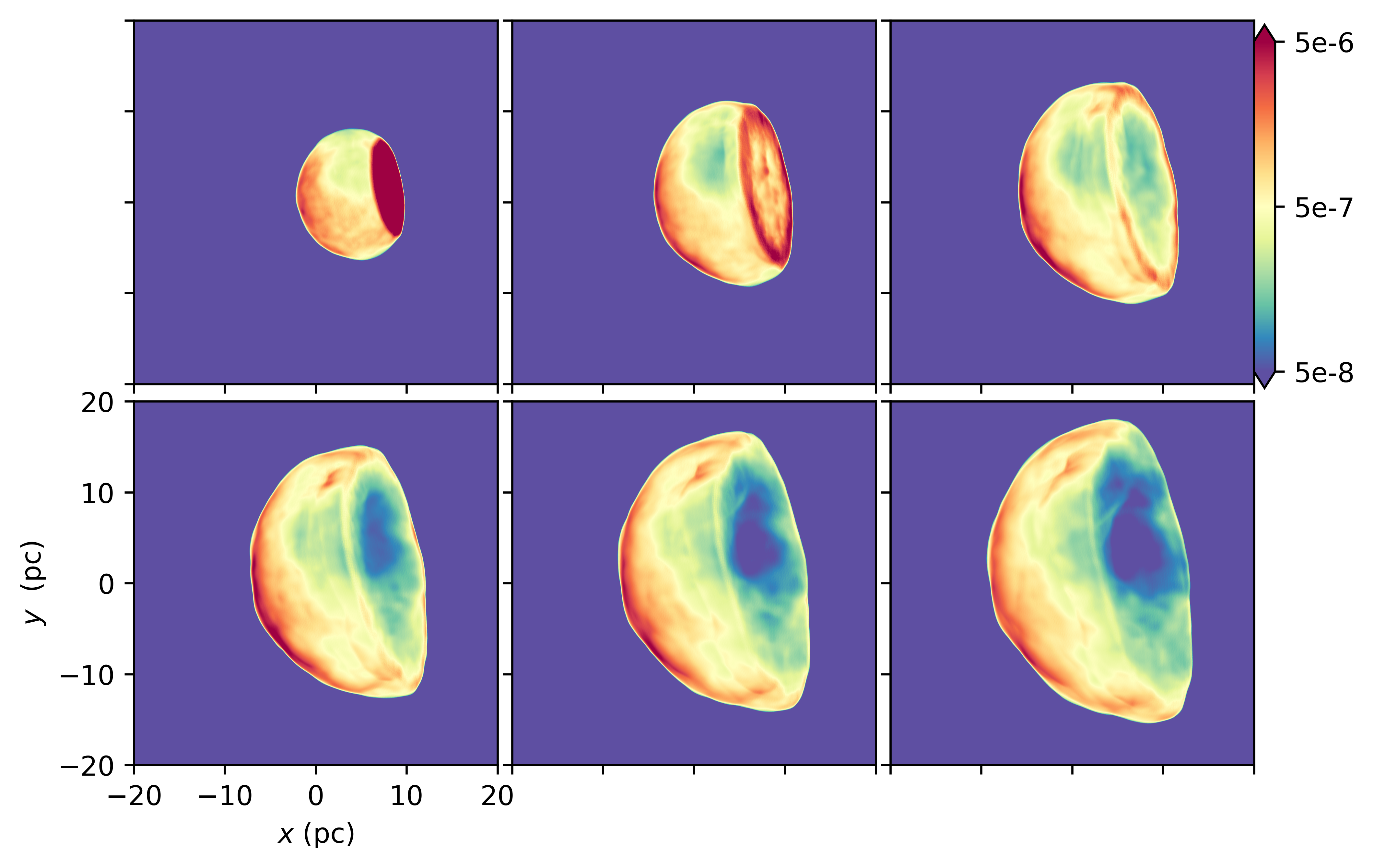}
    \caption{X-ray synthetic map for the case of a rotation of 30 degrees around the x-axis and 20 degrees around the y-axis and for a temporal evolution from 2000 to 12000 yr. The time between panels is 2000 yr. The logarithmic colour scale given by the right bar indicates the X-ray flux (in $\mathrm{erg\ cm^{-2}\ s^{-1} sr^{-1}}$).}
    \label{fig:Fig4}
\end{figure*}

In Figure \ref{fig:Fig4}, we show the temporal evolution of the thermal emission in X-rays. Like the radio emission, the map corresponding to 2000 years of integration shows a bright elliptical zone from the interaction region with the cloud. This elliptical emission becomes an intense emitting ring at 4000 yr. However, unlike the non-thermal emission, we did not observe the presence of the reflected shock at this frequency. The X-ray emission is generally irregular, resulting from the evolution in a medium with a turbulent density distribution. 
From our simulations, we calculate the total X-ray luminosity $L_X$ emitted by the SNR in the range [0.35-8.0] KeV without interstellar absorption, obtaining a value of $2.09\times 10^{35}$ erg s$^{-1}$ (assuming a distance of 3 kpc). According to archival data ({\sc{chandra supernova remnant catalog}}\footnote{\url{https://hea-www.harvard.edu/ChandraSNR/G109.1-01.0}}), this luminosity is slightly less than the observational one ($2.54\times 10^{35}$ erg s$^{-1}$). As mentioned above, we expected our X-ray calculation to underestimate the total $L_X$ due to the ionization equilibrium assumption.

\begin{figure}
	\includegraphics[width=\columnwidth]{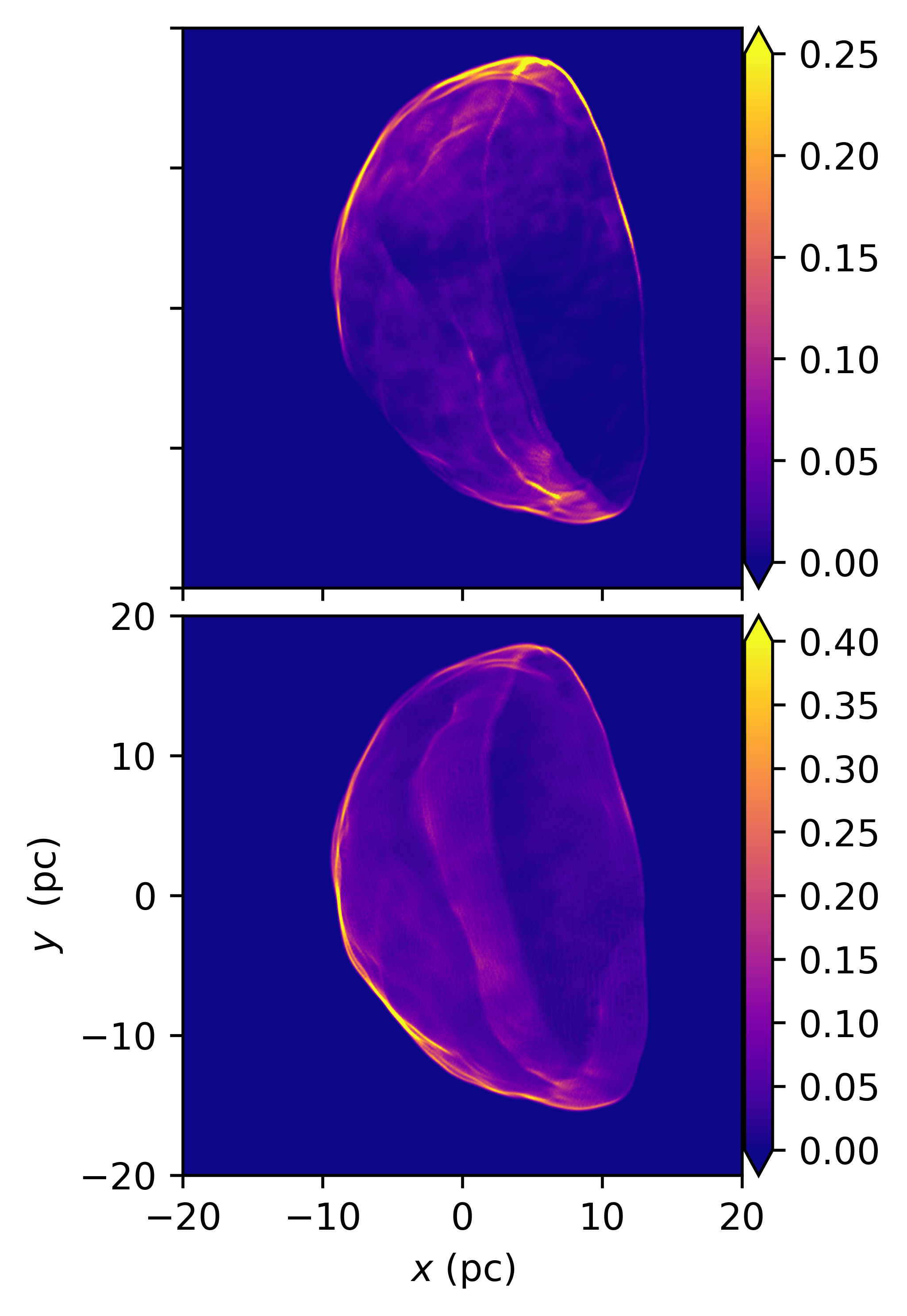}
    \caption{Synthetic map of the synchrotron emission $I$ for a temporal evolution of 12000 yr. The top panel shows the quasi-parallel case, while the bottom panel show the isotropic case. The linear colour scale given by the right bar indicates the synchrotron flux in arbitrary units.}
    \label{fig:Fig5}
\end{figure}

\begin{figure}
	\includegraphics[width=\columnwidth]{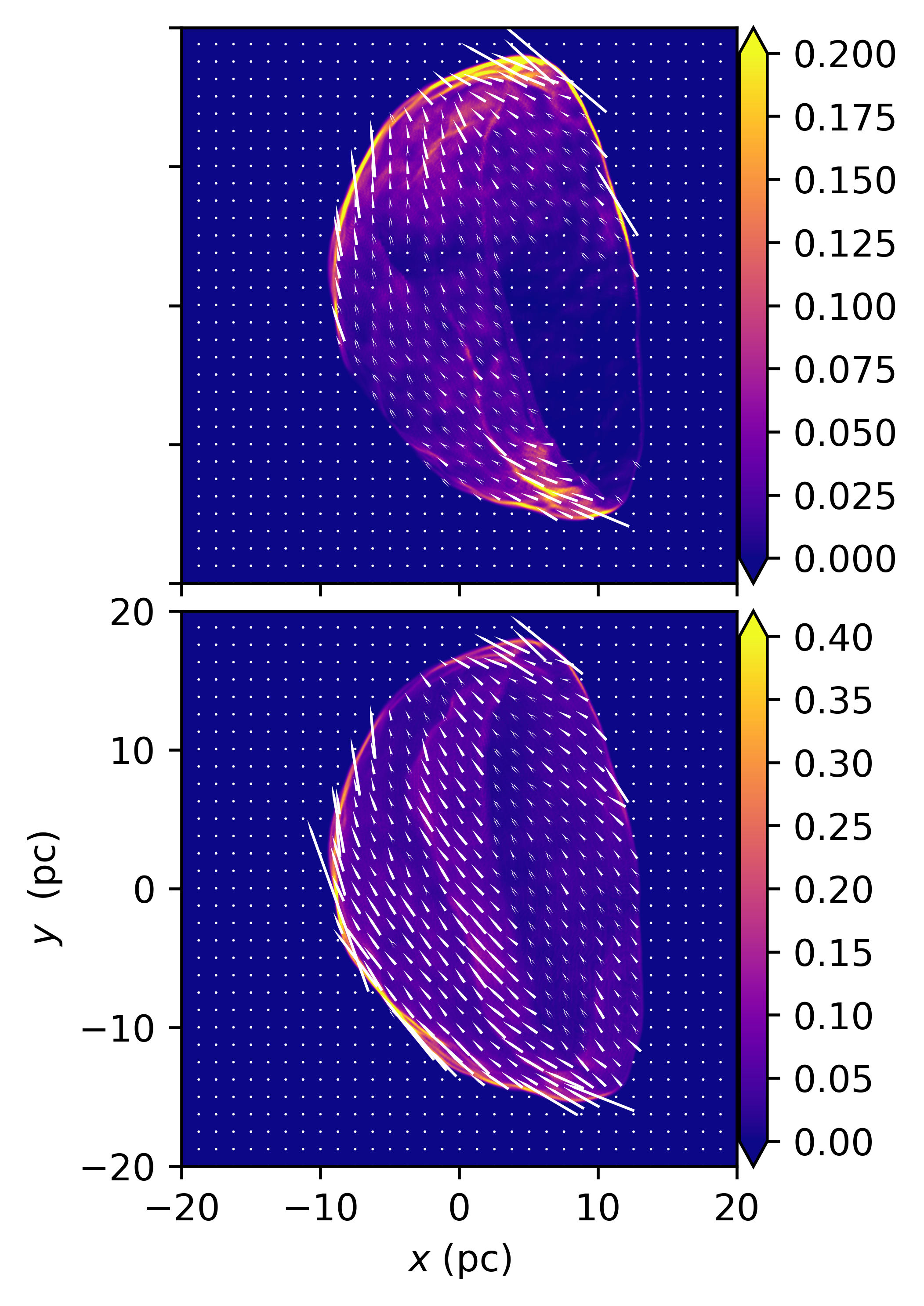}
    \caption{Synthetic map of the linearly polarized emission $I_p$ and for a temporal evolution of 12000 yr. The top panel shows the quasi-parallel case, while the bottom panel show the isotropic case. The linear colour scale given by the right bar indicates the flux in arbitrary units. The white arrows indicate the magnetic field orientations given by $\Phi_B$.}
    \label{fig:Fig6}
\end{figure}

We present the synchrotron  and the linearly polarized radio emissions  in  Figures~\ref{fig:Fig5} and \ref{fig:Fig6}, respectively, for an integration time of  $12000$ yr. Just for comparison,  we show the quasiparallel case maps (top panel in both figures) and the isotropic case maps (bottom panel in both figures). We also show the $\Phi_B$ distribution (white vectors, in Figure \ref{fig:Fig6}) overlaying the polarized intensity emission. For the quasiparallel case, we observed strong radio emission from the top-left region of the remnant. Besides, the  orientation given by $\Phi_B$ is almost orthogonal in the upper half of the remnant. These orientations of the PA and radio emission are in agreement with the observations at 10 GHz carried out by \citet{Tatematsu1987a}. Instead, the map of the isotropic case show strong emissions in the lower part of the remnant. Also, we see PA orientation parallel to the shock front for this case. These last two results are not in agreement with observations.

\begin{figure}
	\includegraphics[width=\columnwidth]{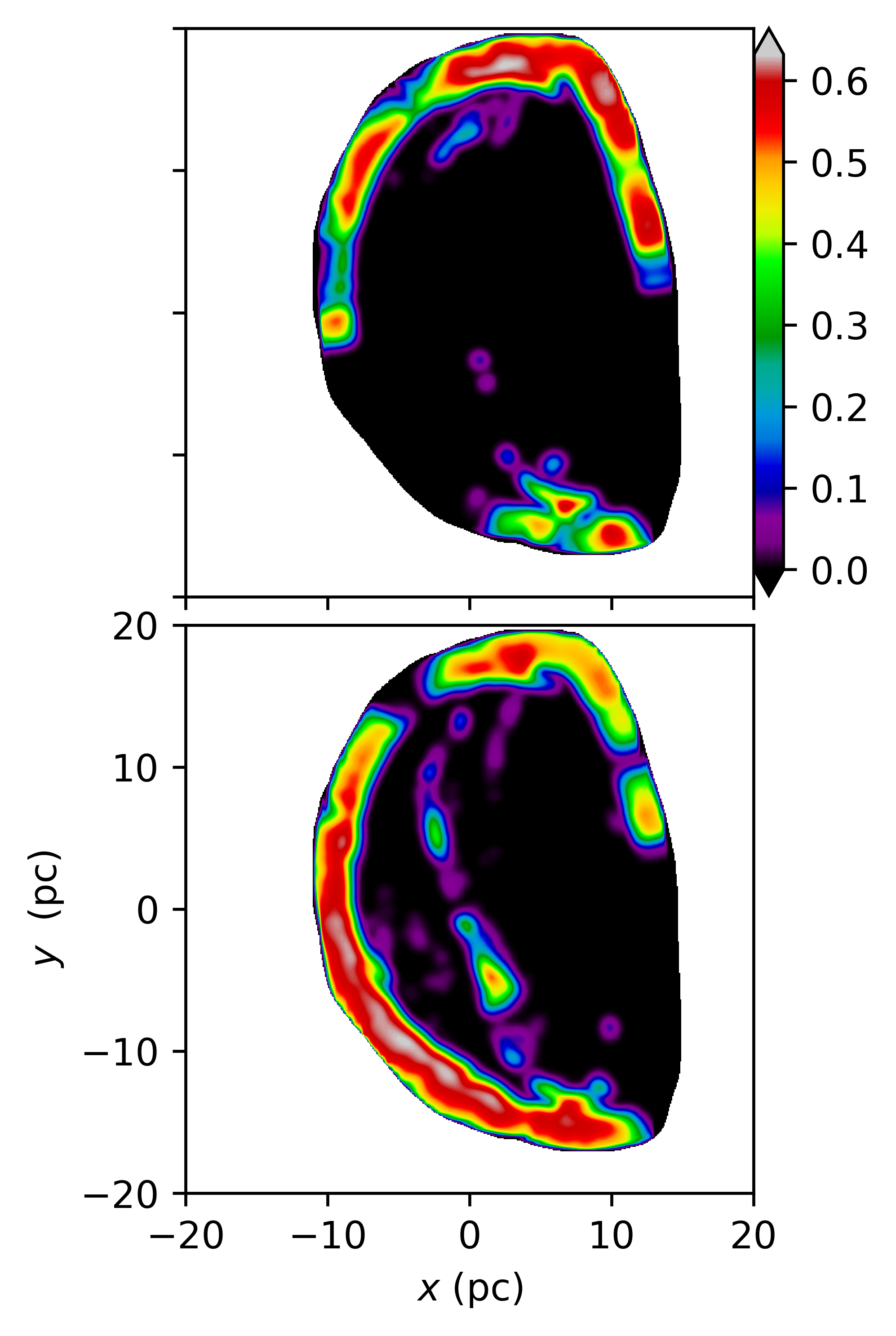}
    \caption{Synthetic map of the polarization fraction for a temporal evolution of 12000 yr. The top panel shows the quasi-parallel case, while the bottom panel show the isotropic case. The fraction is shown with the linear colour scale given by the right bar.}
    \label{fig:Fig7}
\end{figure}

The polarization fraction is 60\% in the brighter regions (see both panels of Fig.~\ref{fig:Fig7}, which show the quasiparallel case in the top and the isotropic in the bottom), twice the reported value \citep{Tatematsu1987a}. This high value is not a surprise because we have imposed a uniform ISMF. Although the SNR evolve into an environment with a turbulent density distribution, this is not enough to significantly reduce the polarization fraction. Therefore, it is necessary to add a random component to the uniform ISMF, such as was shown by \citet{Velazquez2017}.

\section{Conclusions}

This work presents a full 3D MHD model for the SNR CTB109. Considering previous works, we study the evolution, the X-ray and synchrotron emission of a remnant evolving into an ISM with a high-density contrast for simulating the existence of a dense cloud. In addition, we have introduced an inhomogeneous medium by introducing a perturbed medium with a noisy density structure to reproduce more realistic characteristics in our simulation. This perturbed environment is consistent with a turbulent interstellar medium. As a result of our simulations, an SNR with the size ($\sim16$ pc at $1.2\times10^4$ yr) and characteristics of CTB~109 is obtained if the SN explosion occurs in the less dense region, but close to the dense cloud frontier. 

We observe that a double shock structure is developed through the interaction between the ejected material by the SN explosion, the environment and the cloud. The main shock expands into the environment and forms the external shell of the SNR. The secondary shock is due to the rebounce of the main shock against the dense cloud. Finally, we note that in the inner region of the cloud, the shock has practically stopped.

We carried out synthetic X-ray emission maps from our numerical simulations. Structure formation due to inhomogeneities of the environment can be observed. The maximums in the emission correspond to clumps with the highest density, which agrees with the lobe formation in CTB~109. On the other hand, there is no X-ray emission from the inner region of the dense cloud.

Synthetic Stokes parameter maps were performed from our numerical simulations, showing the apparent double shock morphology reported in the observations of this SNR. As we explained above, this double shock structure forms by the interaction of the main shock with the dense cloud. As in the X-ray case, there is no synchrotron emission in the inner region of the cloud. 

We found that the quasiparallel acceleration mechanism explains the synchrotron brightness and PA distributions of SNR CTB109. This result strengthens the conclusion given by \citet{Tatematsu1987a}. These authors say the synchrotron emission is given mainly by an almost uniform ISMF. Therefore, comparing the Stokes parameters emission and the PA distribution is a valuable tool for determining which acceleration mechanism produced the observed morphology in SNRs, such as previous studies have noted \citep[see][]{Aroche2020,moranchel2017, Velazquez2017,Reynoso2013}.

Finally, we found that the X-ray emission in the range [3.5-8.0] keV predicted by our model is in close agreement with the observed value ({\sc{chandra}} archival data).

As far as we know, this is the first time that a full 3D MHD simulation of this object has been carried out. Quite strikingly, our simulation are in good agreement with the observational characteristics of the synchrotron emission from this object. 

\section*{Acknowledgements}

The authors thank the valuable comments and suggestions given by the referee.
The authors acknowledge the financial support for PAPIIT-UNAM grant IA103121. We thank Enrique Palacios (ICN-UNAM) for maintaining the cluster, where the simulations were performed. CHIANTI is a collaborative project involving George Mason University, the Universityof Michigan (USA), University of Cambridge (UK) and NASA Goddard Space Flight Center (USA). A.C.-R. acknowledges support from a DGAPA-UNAM postdoctoral fellowship, and the resources provided by the Miztli supercomputer through the project LANCAD-UNAM-DGTIC-408. We thank an anonymous referee for reading this manuscript and for useful comments that helped us improve this manuscript.

\section*{Data Availability}
The data underlying this article will be shared on reasonable request to the corresponding author.
 



\bibliographystyle{mnras}
\bibliography{references} 








\bsp	
\label{lastpage}
\end{document}